**Developmental tendencies in the Academic Field of Intellectual Property through the Identification of Invisible Colleges**

Guadalupe Palacios-Núñez [1], Gabriel Vélez-Cuartas [2], Juan D. Botero [3].

## Abstract

Intellectual property became a relevant academic cross-disciplinary field in an international context with the demand for the global governance of knowledge. However, the degree of consolidation of cross-disciplinary academic communities is not clear. To determine how closely related are these communities, this paper proposes a mixed methodology to find invisible colleges in the production of intellectual property. Scientific articles from 1994 to 2016 were extracted from Web of Science, taking into account the signature of the agreement on Trade-Related Aspects of Intellectual Property Rights in the early 90's. A total of 1,580 papers were processed through bibliographic coupling network analysis. A special technique was applied, which combines algorithms of community detection and defines a population of articles through thresholds of shared references. To contrast the invisible colleges that emerged with the existence of formal institutional relations, a qualitative tracking of the authors was made with respect to their institutional affiliation, lines of research, and meeting places. Both methods show that the subjects of interest can be grouped into 13 thematic modules related to the intellectual property field. Even though most are related to Law and Economics, there are weak linkages between disciplines which could indicate the construction of a cross-disciplinary field.

[1] Corresponding autor: Instituto de Investigaciones Económicas y Empresariales. Universidad Michoacana de San Nicolás de Hidalgo, Calle Gral. Francisco J. Múgica S/N, C.P.58030, Morelia, Michoacán, México. Orcid ID: 0000-0001-8252-6564. Email: guadalupe_palacios@fevaq.net

[2] Grupo de Investigación Redes y Actores Sociales, Departamento de Sociología, Facultad de Ciencias Sociales y Humanas. Universidad de Antioquia, Calle 70 No. 52-21, Medellín, Colombia. Orcid ID: 0000-0003-2350-4650. Email: gjaime.velez@udea.edu.co

[3] Grupo de Física Atómica y Molecular, Instituto de Física, Facultad de Ciencias Exactas y Naturales. Universidad de Antioquia, Calle 70 No. 52-21, Medellín, Colombia. Orcid ID: 0000-0002-3594-9705. Email: juand.botero@udea.edu.co

**Introduction**

The study of intellectual property (IP) has become important since the signature of the Agreement on Trade-Related Aspects of Intellectual Property Rights (TRIPS) in 1994. Since then, the literature has tackled important debates showing a dynamic field with a diversity of research results. Therefore, scientific publications show that after the signature of TRIPS countries have had to face heterogeneous results and there is no agreement on the costs, benefits, and impacts on innovation, economy, and development. Also, the consequences of implementing IP is still a controversial subject among researchers, which may be due to the empirical evidence on its use which is scarce, ambiguous, and muddled. A general analysis of the state-of-the-art not only shows that it is a controversial subject but a field covering a great diversity of aspects making it difficult to find suitable disciplinary indicators to determine its impacts (Jackson 2003; Lall 2003; Laranja et al. 2008; Montobbio et al. 2015; Rockett 2010).

On the other hand, the number of publications has grown together with bibliometric studies addressing the numbers of articles, of authors, of citations received, of references, and of coauthorships; the most cited journals; the countries of origin of the authors, as well as the publishing sectors, showing that academic institutions are the major contributors (Swain and Panda 2012; Natarajan 2013; Garg and Anjana 2014; Velmurugan and Radhakrishnan 2016). The great diversity of aspects included in the IP field has also raised questions on the composition of its knowledge structure. Garg and Srivastava (2016) made a content analysis of published articles in the Journal of Intellectual Property Rights (JIPR) from 1996 to 2014, through the classification of titles into different subjects. Results allowed them to make inferences on the interests of researchers and to conclude that the field structure is highly fragmented and multidisciplinary (law, science, economics, and management). Although the study of Garg and Srivastava exposes the fragmentation problem, its bibliometric data are restricted to the analysis of articles published in JIPR, which contains literature mainly from India (63,5%) and, in a smaller proportion, international literature (36,5%), the United States being the greatest contributor.

In general terms, it could be said that the aspects addressed in scientific publications describe the general dynamics of the field which include the mention of the existence or non-existence of certain dynamics pertaining to a communitarian structure. Nevertheless, the bibliometric and geographic indicators describe individual and non-communitarian data. The existence of a community is still a hypothesis to be proven while there are no evident bonds between the researchers and the common development of subjects. To observe these communitarian structures that could lead to multiple interconnected or isolated research programs, Derek de Solla Price developed in 1963 the concept of Invisible Colleges (De Solla Price 1963).

Although the term invisible colleges was used for the first time in 1645 by Boyle (Teixeira 2011), its sociology of scientific knowledge dates from the sixties, from the works of Price in 1963. His interest was to reveal the informal communication networks among academics from many institutions, geographically separated, to determine if they constituted significant social groups (Crane 1989). Due to the lack of information on informal communication, its study has been generally based on formal communication structures, identified through publications, and they have been defined as academic groups interacting in a formal and informal way, because they share common interests or scientific objectives, in a specific specialty subject. These subjects are nested in published documents, grouping authors in structured co-citation components according to the shared investigations of interest (Teixeira 2011).

At the time when De Solla Price developed the concept of Invisible College, it was methodologically impossible to achieve the accuracy obtainable today with the relatively recent development of algorithms and applications to detect communities through data already systematized in databases like Web of Science (WoS) or Scopus which did not exist at that time. Subsequent to De Solla Price, the communitarian hypothesis raised in his texts and demonstrated solely from the aggregation of data was observed in an empirical way through tools such as the Main Path Analysis (Hummon and Doreian 1989) or the Algorithmic historiography of Garfield (Garfield et al. 2003).

More recently, a set of methodologies has been refined, not necessarily under the name of invisible colleges, but nonetheless focused on detecting communities. For instance, it is common to find studies of bibliographic coupling based on similarity algorithms (Leydesdorff 2008; Colliander and Ahlgren 2012; Steinert and Hoppe 2017; Ciotti et al. 2016). One of the main problems faced by similarity measures is the significance of results in terms of the detected communities (Meyer-Brötz et al. 2017). It is possible to find some answers through hybrid techniques mixing different similarity measures, modularity procedures, and text- and citation-based analysis (Glänzel and Thijs 2017). Nevertheless, even when invisible colleges could be detected and described by sociological inferences regarding the cognitive coordination and collaboration between researchers (Verspagen and Werker 2003), the similarity measures simply observe statistical relatedness between articles resulting in very fuzzy communities without stable parameters.

The methodologies under the name of invisible colleges recently applied do not differ too much from the bibliographic coupling mentioned above in some cases. For instance, Van Raan (2014) proposes a bibliometric analysis mixing co-word analysis, co-citation, and bibliographic coupling to describe invisible colleges dynamics. Gmür (2003) has applied similarity measures to a co-citation network, comparable to Vogel's (2012) development. Nonetheless, the main difference in these studies dealing with the invisible college concept, are the problems addressed by the authors: institutionalized communities combining co-authorship networks and citation indexes (Kretschermer 1994), social networks of supervisors, students and co-workers (Verspagen and Werker 2003; Brunn and O'Lear 1999), cultural circles as invisible colleges (Chubin 1985); preferential attachment present on collaboration networks from the literature of homophily (Verspagen and Werker 2004), the subject specialty, the scientists as social actors, and the Information Use Environment (IUE) using co-citation and cluster analysis (Zuccala 2006).

Considering the sociological issues involved in the concept of invisible colleges, it is assumed that a community can be described in terms of social cognitive developments. That means the closeness between researchers is built on the basis of shared knowledge, not necessarily on the interactions occurring between them (Luhmann 1996). As more shared texts are read by researchers, more similar should be their thoughts. But these reading dynamics expressed in references co-occur at the same time in as many researchers as there are articles with the same references. The strength of a community does not come from a dyad but from the co-occurrence of these references in groups. Consequently, invisible colleges are formed by researchers that read the same papers and give similar meaning to their concepts, using the same semantic forms to express their findings (Vélez-Cuartas 2013). This affirmation allows groups to be selected according to the number of shared references and well-structured groups, initially dismissing for this purpose a similarity analysis.

Therefore, we observe these communication structures as meaningful information systems that institutionalize knowledge through consistency in well-formed groups, which can be described in terms of bibliographic networks. Consistency is understood in terms of the redundancy of semantic formations supposed in bibliographic coupling. Detecting the natural organization of the network into groups demands comprehensive methods on the structure of information, for which its decomposition into sub-units must be done to discover functional modules (Blondel et al. 2008), which, in this case, serve as invisible colleges interested in a common subject: intellectual property. The subjects of scientific interest can be represented by data connection networks, which are naturally divided into communities or modules forming a structure, susceptible to detection and characterization. The detection of communities requires the partition of the network into communities of densely connected nodes. One method to detect possible divisions in the network structure of communities is modularity, which allows the best division to be found without super-positions and determining the number and size of groups (Newman 2006).

The proper modularity or division of the network topology is that which organizes the communities in such a way that the number of relations among the groups is smaller than the number of relations within the groups. Modularity is considered a network property that measures how well divided the vertices are into communities. Modularity coefficient increases as the sum of the weights between the nodes inside the groups are major than the sum of the weights between the groups compared with all the groups of the network. The magnitude of elements is measured through the determination of how strong is the bond of belonging of an edge to one community or another. High values of modularity correspond to a good division of the network in communities, which are found by optimization techniques or approaching algorithms to find the maximum possible global modularity in big networks (Clauset et al. 2004). If we observe the overall definitions laid out above on invisible colleges, it is possible to relate these modules with invisible colleges.

Since a subject specialty is not in itself an invisible college, but rather invisible college emerge from the density of bonds among communities formed around a specialist subject (Teixeira 2011), we considered it necessary to determine whether an invisible college has been formed with regard to IP. Different from Teixeira´s proposal and previous work based on similarity measures, a non-random or normalized threshold of shared references is established to identify communities through a modularity algorithm. This proposal supposes a community formation based on the stability of communities sharing meaning and not on a relative number of references shared by dyads in a set of articles (i.e. as a function of grouping and not as a function of dyadic behavior in the immediate context of a particular network).

Once a set of thresholds have been chosen, the detection of existing communities and the identification of significant groups in the network is made possible through the thickening of combined groups with the refinement of multilevel algorithms, that perform a local search through an individual iterative movement of the edges in different groups until the modularity increases, producing better results than the single level ones (Noack and Rotta 2009). The partition quality, measured by the modularity, is a scale value between 0 and 1, which measures the density of bonds within the communities compared with the bonds among communities (Blondel et al. 2008). In turn, the modularity of the different networks resulting from cutting edges starting on thresholds are compared to select the best one.

In this context, the objective of the present study is to determine the distribution of communities around the subject of IP based on scientific articles, through the identification and characterization of invisible colleges by the method of bibliographic coupling and detection of clusters through modularity algorithms. The resulting clusters are considered semantic communities, i.e. texts, which independent of the impact of an author and his/her citation of its neighbors, have a common referencing pattern sharing meaning, that is, texts that may share common subjects because their authors are reading basically the same literature. We have called these linkages Networks of Meaning (Vélez-Cuartas 2013). Furthermore, if the resulting modules commonly share productive authors as leaders, use the same terms to denominate their subjects and exhibit a stable group structure of bibliographic coupling, then the community hypothesis can be proved. Thereby, it is possible to determine the diversity of problems addressed in a knowledge field by identifying thematic groups. However, identifying clusters or significant groups through shared references poses the basic problem of establishing a significance threshold. In this article we propose a solution to this problem.

### Data and Methods

To determine the invisible colleges concerning IP and to analyze the existence as well as the characterization of invisible colleges, a search by subject was made in the main database of Web of Science through the equation "(innovation OR knowledge) AND (legislation OR law OR rights) OR (intellectual property)." The search was limited to scientific articles published between 1994 and 2016, in research areas where the majority of publications is focused: business economics, public management, and government laws. The result was 1,580 articles, a total of 94,743 references were extracted from them. The characterization of data is shown in Table 1.

Table 1. Data of scientific articles on the subject of intellectual property.

|  | Articles | Number of References | Number of Authors | Authors Affiliations | Citations | Number of Countries |
|---|---|---|---|---|---|---|
| Total | 1,580 | 94,743 | 2,269 | 2,409 | 27,924 | 59 |
| Average |  | 59.96 | 1.43 | 1.52 | 17.67 |  |

Source: Compiled by authors with Web of Science data (1994-2016).

Table 2 shows the ranking of countries with greater production volume on the subject, obtained from the location of those institutions to which the authors are affiliated. Countries appearing in the table match with the geographic location of the 10 most influential authors (USA) and with the physical space where researchers in the subject meet (USA, England, China and Spain).

Table 2. Ranking of countries with greatest production of articles on the subject of intellectual property.

| Rank | Country | Freq |
|------|---------|------|
| 1 | USA | 998 |
| 2 | Italy | 227 |
| 3 | England | 225 |
| 4 | Germany | 210 |
| 5 | China | 140 |
| 6 | France | 139 |
| 7 | Taiwan | 124 |
| 8 | Netherlands | 122 |
| 9 | Japan | 107 |
| 10 | Spain | 96 |

Source: Compiled by authors with Web of Science data (1994-2016).

A relational matrix of correspondence between articles and references was constructed using the isi.exe software (available at http://www.leydesdorff.net). This software generated a list of articles associated with the total of their references, which served as an input to construct a two-mode network in Pajek (Batagelj and Mrvar 1998). Pajek made possible the construction of a network defined as shared references among a set of articles (Garfield et al. 2003; Leydesdorff 1998). The identification of the number of shared references among articles allowed significant relations to be observed which were calculated based on a threshold of shared references depending on the structure of groups that emerged.

To find the emergence of invisible colleges, we used the VOS Clustering method (Waltman et al. 2010) available in Pajek. The main goal of the algorithm is to minimize the distance between nodes weighted by the association strength between them (Van Eck and Waltman 2009). When this method is applied, the proper selection of the resolution parameter $r$ is significantly important because it is related to the number of detected communities. Higher resolution parameters were found to produce a higher number of clusters (NClust) and vice versa, and a resolution parameter of r=1 produces the same results of the modularity function proposed by Newman and Girvan (2004). To test the quality of the results, the VOS algorithm in Pajek provides a quality factor (VOSQ), which is a number between 0 and 1. The closer to 1 the better the quality of grouping.

However, it is not possible to establish mathematically a unique significant threshold to observe the similarity between two articles by shared references, because significance varies among disciplines and subjects. Nevertheless, through an exploratory method of testing different thresholds and observing the emerging groups of articles, it was possible to determine a significance threshold. Accordingly, figure 1 shows the variations in the values obtained for VOSQ and the NClust, as a result of using the algorithm of VOS clustering and removing shared relations. Three resolution parameters 0.5, 1.0, and 1.5 were applied based on previous work to identify communities in Astronomy and Astrophysics (Glanzel & Thijs, 2017). The results show that regardless of the resolution parameter (0.5, 1.0 or 1.5), from two shared relations the value of VOS Quality increases as the low values of shared relations are removed. In the same way, despite the small fluctuations, an increase in the number of clusters is observed as a greater number of shared references are eliminated.

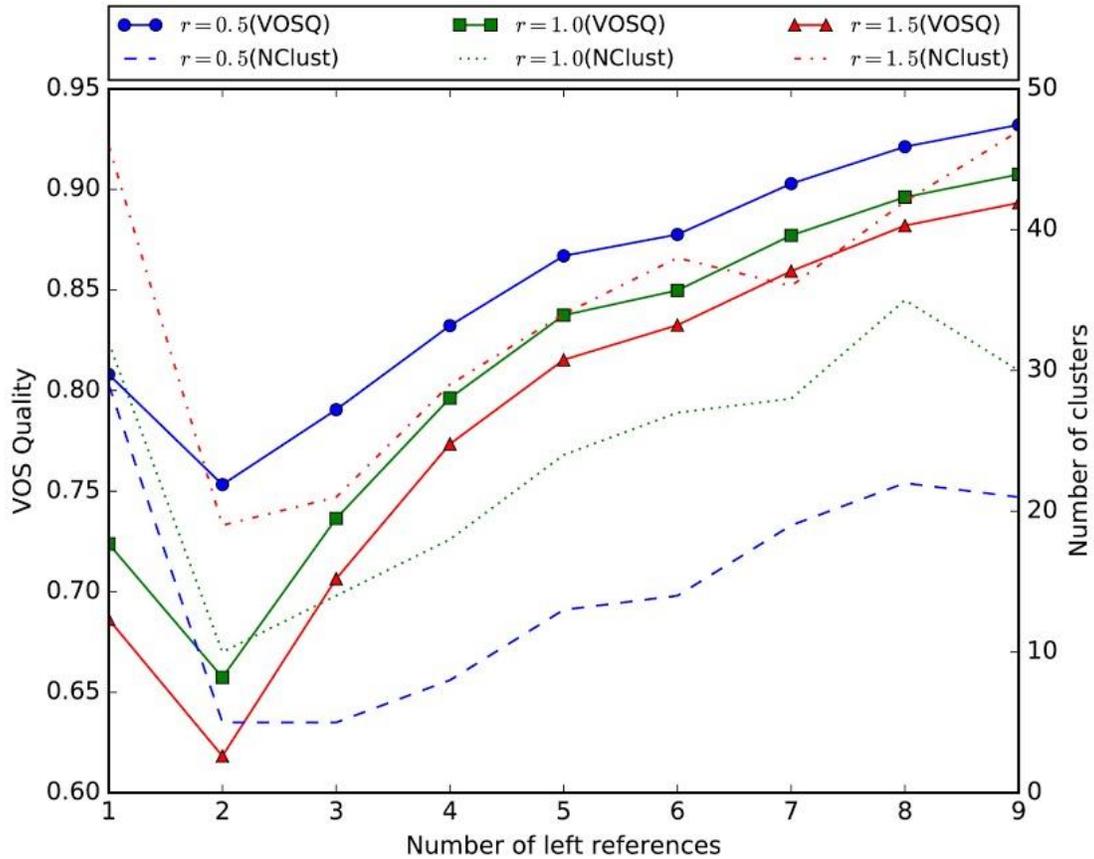

**Fig 1** Detection of communities through shared references. The graph depicts the values of the VOS quality and the number of clusters obtained for different values of the resolution parameter r=0.5, 1.0, 1.5.

To select the best threshold for the shared references in the process of clustering, we took into account that the VOS quality number must be the higher. In addition, it was considered that the fewer the number of clusters, the easier it is for them to be described in terms of cognitive communities. Furthermore, in order not to affect the number of references included in the network, the smallest possible number of links were removed, i.e. it was considered that this threshold must be the lowest possible value.

The identification of the most suitable value for the threshold in figure 1 was most important because the behavior of the curves was not smooth nor uniform, showing its implication for the process of grouping. To obtain a more accurate value for the selection of the threshold, an analysis was made of the slope value in each section of the curve and the results are shown in figure 2. It is apparent that in section number 5, a local minimum appears in the value of slopes of the VOS Quality. Assigning this minimum value, the removal of shared relations does not generate a significant increase in the VOS quality value, and it also satisfies the conditions of securing the lowest number of clusters and links. Based on that, it can be stated that the most significant threshold for shared relations was five from the clarity in the definition of subgroups that was obtained.

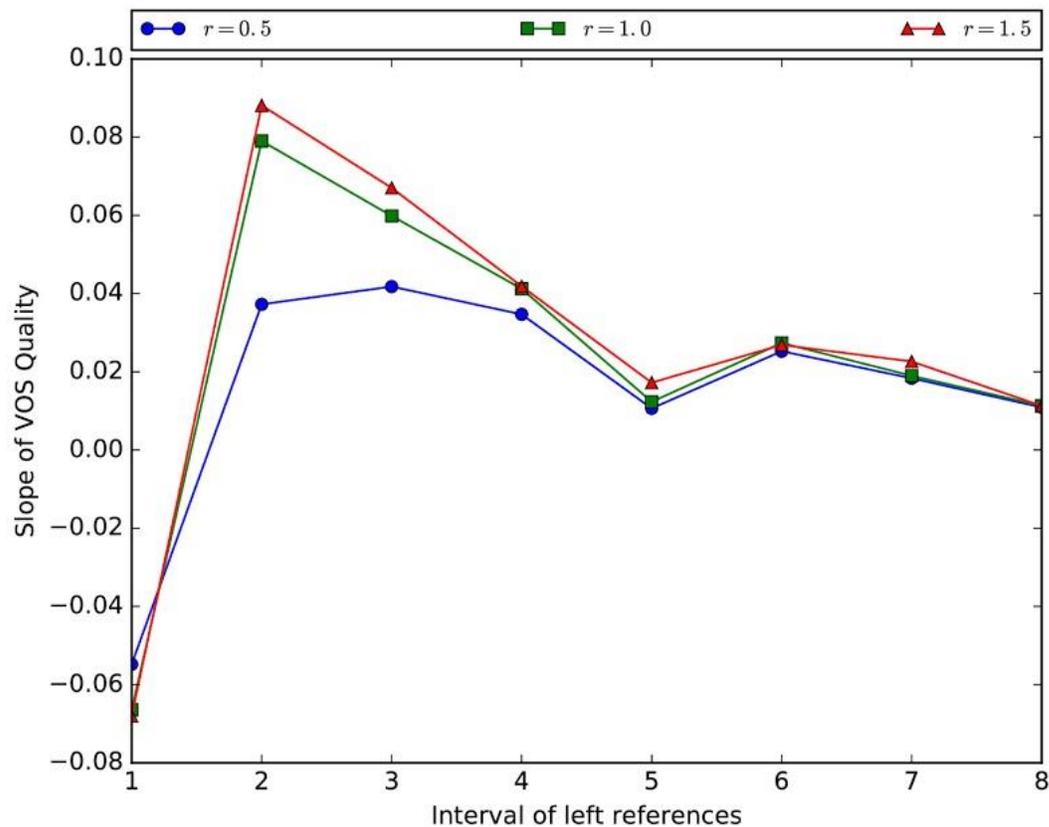

**Fig 2** Identification of optimal modularity to detect communities. The figure represents the slopes of the curves of VOS quality presented in Fig 1.

Consequently, the communities were constructed by identifying those articles sharing at least five joint references and by removing relations with a lower threshold. After the greater component was extracted, a total of 1,267 articles remained and the other 313 articles that did not share five or more common references were eliminated as having no significant relations. The VOS clustering algorithm was then applied whose modularity method is based on the optimization of the quality function. A 0.5 refinement was applied on the partition obtained in the last level through the function "Multi-Level Coarsening + single Refinement" available in Pajek. As a result, a total of 13 communities was obtained (Fig. 3) with VOS Quality= 0,866974. It is a value close to one, which indicates that the refinement gathered the edges in groups in such a way that there is a greater number of edges within the groups than among the groups (Clauset et al. 2004).

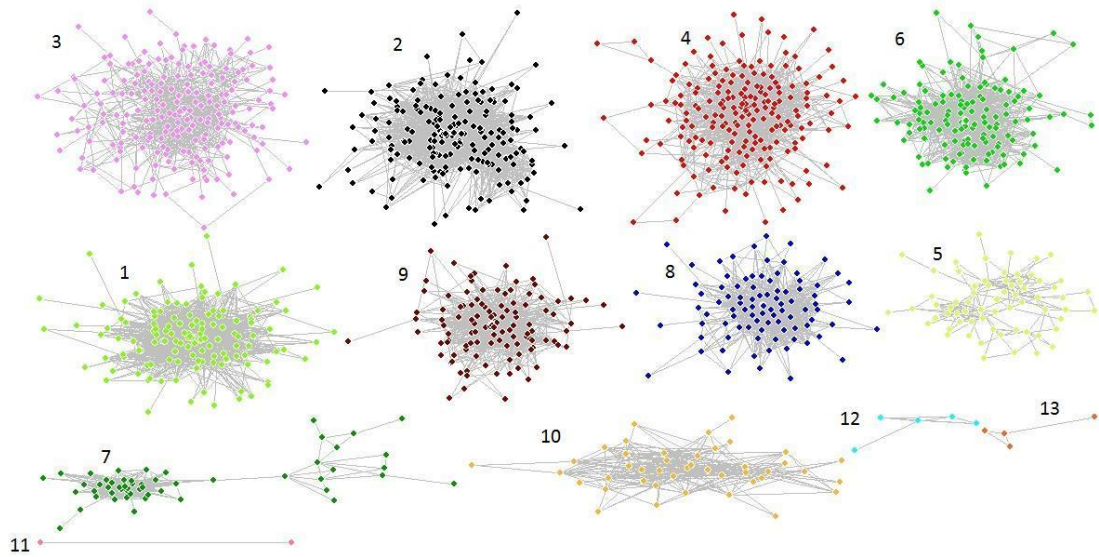

**Fig 3** Thematic modules identified by the modularity. The 13 communities detected after the clustering process and listed in Table 3, are presented as a set of networks with links representing shared references between the articles.

Finally, once the 13 communities were obtained, thematic modules were built through a semantic association of the titles of the articles contained in each one of them. After that, each community was labeled with a title integrating general subjects of interests (Table 3).

Table 3. Thematic modules.

| Community | General subject | Number of articles | % of 1,267 |
|:---:|:---|:---:|:---:|
| 3 | Schumpeter legacy, innovation capabilities, knowledge flows, R&D spillovers and spatial proximity. | 204 | 16,1 |
| 2 | The impact of IPR on foreign direct investment in developing countries and economic growth. | 200 | 15,8 |
| 4 | Patent information for quantitative estimation of technological impact. | 199 | 15,7 |
| 6 | Appropriability strategies, Open Source, markets for technology and Intellectual Property. | 143 | 11,3 |
| 1 | Pirates, innovation by imitation, patent trolls and rethinking IP and the commons. | 136 | 10,7 |
| 9 | Biotechnology industry, university-industry collaborations, academic and industry patents. | 119 | 9,4 |
| 8 | Infringement and litigation of patents, patent thicket, patent pooling, benefits and costs of strong patent protection. | 92 | 7,3 |

| 5 | Corporate innovation and venture capital. | 66 | 5,2 |
|---|---|---|---|
| 7 | Patent licensing, knowledge flows and spillovers. | 54 | 4,3 |
| 10 | Models of technological change, R&D, regulations, policy, and transfer of climate change mitigation technologies or eco innovations. | 43 | 3,4 |
| 12 | Pharmaceutical innovation and drug patenting | 5 | 0,4 |
| 13 | Technological convergence and patent indicators | 4 | 0.3 |
| 11 | Innovation in China | 2 | 0,2 |

Source: Compiled by authors with Web of Science data (1994-2016).

**Results and discussion**

The result allows inferences to be made on the interests of researchers and as can be seen, these point towards specific programs of research within the knowledge-based economy, conforming invisible colleges. The analysis made by Garg and Srivastava (2016) on the structure of knowledge had already shown Economics as one of several disciplines studying IP, but it was not evident that it is a governing axis of the research subjects, that is why these authors concluded that it is a fragmented field. The evidence of our study demonstrates that Economics constitutes a cross-sectional subject for the global scientific production in IP in contrast to what was found previously.

The 13 communities detected by the modularity algorithm, once classified into general interest subjects, show that 47,6% of the edges are concentrated in economics subjects derived from Schumpeterian and Neoclassical theories, which can be observed with the merging of communities 3, 2 and 4. On the other hand, communities 6, 1, 9, and 8, which constitute 38,7%, are focused on the costs and benefits of IP but mixing economics and law in their studies. Small communities 5, 7, and 10, show that 12,9% remain within the scope of economics, but they add subjects related to business, eco-innovations, and public policies. The small satellites (12, 13, and 11) constitute 0,9% and, although their focus is on case studies, these are also related to economics subjects and are therefore connected to the largest communities through citation to the most influential authors.

The fact that all modules display a relationship with economics, matches the fact that the most influential authors are affiliated to economics departments (Table 4). Furthermore, the intersection of the semantic sense and the co-citation, makes visible the fact that, although there is a diversity of subjects, there is a mainstream of authors who constitute invisible colleges, and whose common interest is the knowledge-based economy. Although this common interest is not entirely explicit in the words used, the revision of the semantic content of the modules shows the predominance of a concern for the relation between variables of the economy with IP and knowledge.

Economics appears as the main discipline, which can be explained by the developments in the theoretical and methodological framework for studying the knowledge-based economy. The field is embedded in neoclassical economy and Schumpeterian models, which equip Intellectual Property Rights (IPRs) with a crucial role to provide incentives for the private creation of knowledge. In this direction, the innovation systems approach emphasizes the importance of institutions to mediate the main activities of the knowledge-based economy, which are directly based on production, distribution, and use of knowledge to create value. The IPRs, therefore, became relevant as a result of the globalization of innovation and the demand for global knowledge governance. This demand was met through the TRIPS, which offered a framework of meta-regulation of knowledge, and consequently, the studies dedicated to adding empirical evidence of their efficiency have proliferated (OCDE 1996; Harris 2001; Laranja et al. 2008; Montobbio et al. 2015). All these issues are embedded in the discussions alluded to in the thematic modules in Table 3, close to the field of economics.

The titles of articles show that the authors have a common interest in obtaining empirical evidence on the effectiveness of the patents system in the creation of incentives for innovation, i.e. to study the relation of IPRs with the production of knowledge in the private and academic sector. Additionally, other analyses carried out concern the costs and benefits of strengthening the IPRs and their relation to increases in economic variables such as economic growth, R&D investment, knowledge spillovers, knowledge diffusion, innovation rates in different technological sectors, R&D outputs, effects of geographically located knowledge, endogenous growth, foreign direct investment, royalties, markets of technology, and technological change. Since citations are the main indicator of the influence of a scientist in their specialty (Teixeira 2011), citations were counted to rank the most influential authors on the subject (Table 4).

Table 4. Most influential authors.

| Rank | Most influential authors | Institutional affiliations and department | Number of citations | Citing communities |
|---|---|---|---|---|
| 1 | Griliches Z, 1990, J ECON LIT, V28, P1661 | Harvard University Department of Economics | 254 | 2, 3, 4, 6, 10 |
| 2 | Hall BH, 2001, RAND J ECON, V32, P101 | University of California, Berkeley Department of Economics | 189 | 3, 5, 8 |
| 3 | Levin R.C., 1987, BROOKINGS PAPERS EC, V1987, P783 | Yale University Department of Economics | 188 | 1, 2, 3, 4, 5, 6, 8, 10 |
| 4 | Heller MA, 1998, SCIENCE, V280, P698 | University of Michigan Law School | 161 | 1, 6, 8, 9 |
| 5 | Cohen WM, 1990, ADMIN SCI QUART, V35, P128 | Carnegie Mellon University Department of Economics | 150 | 3, 4, 6, 9 |
| 6 | Teece DJ, 1986, RES POLICY, V15, P285 | University of California, Berkeley Institute for Business Innovation | 146 | 4, 6 |
| 7 | Hall BH, 2005, RAND J ECON, V36, P16 | University of California, Berkeley Department of Economics | 144 | 3, 4, 5, 6 |
| 8 | Jaffe AB, 1993, Q J ECON, V108, P577 | Harvard University Department of Economics | 139 | 3, 4, 5, 9, 10 |
| 9 | Trajtenberg M, 1990, RAND J ECON, V21, P172 | Tel-Aviv University Department of Economics | 130 | 3, 4, 5, 9 |
| 10 | Merges RP, 1990, COLUMBIA LAW REV, V90, P839 | Boston University School of Law | 127 | 1, 6, 8, 9 |

Source: Compiled by authors with Web of Science data (1994-2016).

Identification of the most influential authors, their institutional affiliations, and their lines of research corroborate the fact that the main interest of communities is within economics (80%) and, in a smaller proportion, in law (20%). Likewise, Table 4 shows that the majority of the most influential authors are affiliated to North American universities: Griliches (Harvard University), Hall (University of California, Berkeley), Levin (Yale University), Cohen (Carnegie Mellon University), Teece (University of California, Berkeley), Jaffe (Harvard University), Heller (University of Michigan) and Merges (Boston University), with the exception of Trajtenberg (Tel Aviv University).

What is evident is a concern for several aspects of the knowledge-based economy in the lines of research of the most influential authors, where the law is applied considering that IP is a fundamental institution for the global governance of knowledge. The main lines of research of these authors are the economy of technological change, empirical studies of innovations diffusion, R&D outputs, patents and econometrics, Science, Technology and Innovation Policy, law theory, determinants of innovating activity, benefits of technological innovation, dynamic capacities and strategic management, defense of competition, law of patents, and IP in the new technological era.

The results show invisible colleges in the structure of the communities, in the sense that was theoretically put forward by De Solla Price (1963) for their identification. He proposed a structure where there are prestigious authors and, around them, other researchers who are mentioned in the text set corresponding to the invisible colleges. As a result, IP is constituted as a science topic, at the same time that these most influential authors appear in different colleges where different subjects are being questioned around this central axis, as can be observed in the last column of Table 4. Even though in communities 7, 11, 12, and 13, these authors do not appear, they share the semantic presence of terms related to IP. This confirms the existence of interrelated emergent fields with a main current of concerns, but with weak bonds due to not sharing references with the 'mainstream' literature. This can be interpreted as other aspects of the IP discussions that grant an emergent character to these subjects within the semantic field of this object of study.

**Conclusions**

This paper proposes a new procedure to establish invisible colleges from the semantic structure emerging from the text sets sharing a specific threshold of references. This methodology has turned out to be useful to identify the conformation of invisible colleges in specific subjects within a general thematic field like IP. The methodology has also allowed the identification of the interconnection between these isolated invisible colleges, conforming denser and more complex groups of subjects associated with a general knowledge field. This could determine the direction of a specific subject of science and at the same time identify interrelated subjects through visible authors who constitute the field according to the definition of invisible colleges of De Solla Price. In addition, the methodology also allowed identifying emergent subjects that are not interrelated via the most prestigious authors in the field, but connect to the main thematic field as shown by common concepts shared from different points of view.

The connections between the 13 communities obtained through the modularity algorithm and its corresponding semantic association with general interest subjects, reveal the existence of invisible colleges whose specific research programs are aimed towards the knowledge-based economy. This is inferred from a revision of the semantic sense of the modules, which show that the main concern is the relation between variables of the economy and intellectual property and knowledge. The economy appears as the main discipline because the theoretical and methodological framework for studying the knowledge-based economy is embedded in the neoclassical economy and Schumpeterian models, which equip Intellectual Property Rights (IPRs) with a crucial role to provide incentives for the private creation of knowledge.

This finding leads us to the conclusion that research on intellectual property is not highly fragmented as was shown by Garg and Srivastava (2016) in their paper on the structure of knowledge in intellectual property rights, but rather the knowledge-based economy serves as a governing axis. Although it is true that there are multidisciplinary contributions on the subject, with our method it was possible to observe a clear predominance of economics. The content analysis and classification method used by Garg and Srivastava (2016) showed the diversity of aspects included in the IP field. However, it was unable to identify the general interest responsible for the constitution of invisible colleges. In this sense, the methodology applied in our work was more accurate and suitable for this purpose.

From the methodological point of view, the modularity method was able to identify the existence of invisible colleges, detected from references in formal scientific communications. Unlike other methodologies (Hummon and Doreian 1989; Garg and Srivastava 2016; Garfield et al. 2003), the identification of invisible colleges from the analysis of co-references (bibliographic coupling) and the establishment of thresholds for shared references to identify communities, allows not only the constitution of a 'mainstream' in which there are several thematic related communities, but also emergent subjects within a knowledge field, in this case, IP. Different from the similarity measures applied by other researchers (Leydesdorff, 2008; Colliander & Ahlgren, 2012; Steinert & Hoppe, 2017; Ciotti et al, 2016), identifying thresholds of shared references without previous normalization, enabled the discovery of nuclear communities in the development of different subjects in a knowledge field and not only sectors in the network with higher similarities.

Finally, ours was a global approach to knowledge structure, and the resulting communities made it possible to classify subgroups of subjects representing the interests of the researchers. Furthermore, the study evidences the need to research the characterization of communities in a deeper way, for which future studies will be necessary to determine the degree of cohesion of the groups, the degree of diversification in the subject, the centrality of authors and to establish if intellectual property is being constituted as an independent research field of economics, public management, and law or if it is still a fragmented specialty subject. We also need to find out if the fragmentation displayed by the field indicates the emergence of subjects within the specialty.

## Acknowlegements

We are sincerely grateful to Jane Russell and Maria Victoria Guzmán for their meticulous comments and suggestions, which were very significant and helpful to improve our work. We also like to mention that this work was done thanks to supporting of Colciencias through a doctoral scholarship for Juan D. Botero (Program No. 727).